\documentstyle[emulateapj,psfig,natbib,amsfonts,amsmath]{article}
\def\swift{{\it Swift}}
\def\nod{\nodata}
\def\grb{GRB\,051111}

\def\ociw{1}
\def\prince{2}
\def\hubble{3}
\def\pom{4}
\def\psu{5}
\def\cit{6}
\def\keck{7}

\begin{document}

\title{Fine-Structure \ion{Fe}{2} and \ion{Si}{2} Absorption in the
Spectrum of GRB\,051111: Implications for the Burst Environment}

\author{
E.~Berger\altaffilmark{\ociw,}\altaffilmark{\prince,}\altaffilmark{\hubble},
B.~E.~Penprase\altaffilmark{\pom},
D.~B.~Fox\altaffilmark{\psu}, 
S.~R.~Kulkarni\altaffilmark{\cit}, 
G.~Hill\altaffilmark{\keck},
B. Schaefer\altaffilmark{\keck}, 
and M. Reed\altaffilmark{\keck}
}

\altaffiltext{\ociw}{Observatories of the Carnegie Institution
of Washington, 813 Santa Barbara Street, Pasadena, CA 91101}
 
\altaffiltext{\prince}{Princeton University Observatory,
Peyton Hall, Ivy Lane, Princeton, NJ 08544}
 
\altaffiltext{\hubble}{Hubble Fellow}

\altaffiltext{\pom}{Pomona College Department of Physics and Astronomy,
610 N. College Avenue, Claremont, CA}

\altaffiltext{\psu}{Department of Astronomy and Astrophysics,
Pennsylvania State University, 525 Davey Laboratory, University
Park, PA 16802}

\altaffiltext{\cit}{Division of Physics, Mathematics and Astronomy,
105-24, California Institute of Technology, Pasadena, CA 91125}

\altaffiltext{\keck}{W.~M.~Keck Observatory, 65-1120 Mamalahoa 
Highway, Kamuela, HI 96743}

\begin{abstract} We present an analysis of fine-structure transitions
of \ion{Fe}{2} and \ion{Si}{2} detected in a high-resolution optical
spectrum of the afterglow of \grb\ ($z=1.54948$).  The fine-structure
absorption features arising from \ion{Fe}{2}* to \ion{Fe}{2}****, as
well as \ion{Si}{2}*, are confined to a narrow velocity structure
extending over $\pm 30$ km s$^{-1}$, which we interpret as the burst
local environment, most likely a star forming region.  We investigate
two scenarios for the excitation of the fine-structure levels by
collisions with electrons and radiative pumping by an infra-red or
ultra-violet radiation field produced by intense star formation in the
GRB environment, or by the GRB afterglow itself.  We find that the
conditions required for collisional excitation of \ion{Fe}{2}
fine-structure states cannot be easily reconciled with the relatively
weak \ion{Si}{2}* absorption.  Radiative pumping by either IR or UV
emission requires $>10^3$ massive hot OB stars within a compact
star-forming region a few pc in size, and in the case of IR pumping a
large dust content.  On the other hand, it is possible that the GRB
itself provides the source of IR and/or UV radiation, in which case we
estimate that the excitation takes place at a distance of $\sim 10-20$
pc from the burst.  Detailed radiative transfer calculations are
required in order to verify that excitation of the low-ionization
fine-structure states is possible given the intense UV flux from the
burst.  Still, it is clear that GRB absorption spectroscopy can
provide direct information on the mode and conditions of star
formation at high redshift.  \end{abstract}
 

\section{Introduction}
\label{sec:intro}

Understanding the physical conditions in star forming regions at high
redshift plays a crucial role in our ability to trace the evolution of
star formation and the associated production of metals.
Unfortunately, it is exceedingly difficult to probe these regions
using background quasars because of their compact size, and possible
dust extinction.  Absorption studies of bright star forming galaxies
(e.g., the lensed Lyman break galaxy MS 1512$-$cB58; \citealt{prs+02})
provide greater promise because they probe the regions where star
formation is taking place, but even in these rare cases information is
only available on the {\it integrated} properties and individual
regions cannot be probed.

Gamma-ray bursts (GRBs), on the other hand, are known to be the end
product of massive stars, and as a result their bright afterglows can
be used to trace the interstellar medium of their host galaxies.  More
importantly, since GRB progenitors are short-lived, they are likely to
be buried within star-forming regions undergoing active star formation
\citep{bkd02}.  As a result, GRBs may provide an ideal direct probe of
the physical conditions in high redshift star forming regions.  Over
the last several years, and particularly with the advent of the {\it
Swift} satellite, several GRB absorption spectra have been obtained
revealing a large fraction of objects with unusually large neutral
hydrogen column densities even for damped Ly$\alpha$ (DLA) systems,
$N\gtrsim 10^{22}$ cm$^{-2}$ (e.g., \citealt{vel+04,bpc+05}).  In
addition, it has been noted in several cases that the column densities
of non-refractory elements (e.g., Zn) are higher than in
QSO-DLAs, with a depletion pattern suggestive of a large dust content
\citep{sf04}.  These properties suggest that at least some GRB-DLAs
are physically different from QSO-DLAs, and that they may represent
individual star-forming regions with at least a modest dust
obscuration.

Given the possible association with star-forming regions, several
authors have also investigated the impact of the GRB prompt and
afterglow emission on the local environment, particularly in the
context of dust destruction, photo-ionization, and dissociation and
excitation of $H_2$ molecules \citep{dh02}.  The basic conclusion of
these studies is that bright afterglows are capable of significantly
modifying their local environment on a scale of about $10^{19}$ cm.
The detection of such effects can therefore provide a unique
diagnostic of the immediate environment of the burst, and in addition
may address the issue of relatively low dust extinction in GRB optical
afterglows.

In this {\it Letter} we present a high-resolution absorption spectrum
of \grb, which reveals strong absorption features from \ion{Fe}{2} and
\ion{Si}{2} fine-structure states at a redshift $z=1.54948$.  This is
the first example of an extra-galactic sight line that reveals the
full range of \ion{Fe}{2} ground-level fine-structure transitions,
suggesting that their excitation is intimately related to the burst
environment or the burst itself.  We investigate both possibilities
and their implications in the following sections.

\section{Observations}
\label{sec:obs}

\grb\ was detected by \swift\ on 2005 November 11.250 UT
\citep{gcn4260}, and the optical afterglow was discovered with the
ROTSE-IIIb robotic telescope 27 s after the burst with an unfiltered
magnitude of about 13 \citep{gcn4247}.  Follow-up observations in the
first hour after the burst indicate an optical flux evolution,
$F_\nu(R)\approx 280(t/1\,{\rm hr})^{-0.9}$ $\mu$Jy \citep{gcn4267}.

Spectroscopic observations of \grb\ were initiated approximately 1 hr
after the burst, using the High Resolution Echelle Spectrometer
(HIRES) mounted on the Keck I 10-m telescope \citep{gcn4255}.  A total
of 5400 s were obtained in three exposures using a $0.86''$ wide slit.
The wavelength range is $4200-8400$ \AA.  The spectra were reduced
using the Makee pipeline routines (Version 4.0.1 of May 2005), which
includes optimal extraction of orders, sky subtraction, and wavelength
calibration from Th-Ar arc lamp exposures, including a heliocentric
velocity correction.  The orders within individual frames were traced
using a median combined total of the exposures, and atmospheric
absorption features were removed with the Makee pipeline.  A final
resampling of the spectrum and continuum fitting was performed using
the IRAF task {\tt continuum}.

The spectrum reveals strong absorption features at a redshift,
$z=1.54948\pm 0.00001$, which we interpret to arise in the host galaxy
(see also \citealt{gcn4271}\footnotemark\footnotetext{A GRB
Coordinates Network Circular by \citet{gcn4271} provides a cursory
list of some properties of the host galaxy absorption system,
including the detection of fine-structure lines.}).  A full analysis
of the host galaxy and intervening absorption systems is presented in
a companion paper (Penprase et al. 2005).  Here we focus on the
detection of \ion{Fe}{2} and \ion{Si}{2} fine-structure transitions.

\section{\ion{Fe}{2} and \ion{Si}{2} Fine-Structure Transitions} 
\label{sec:abs}

\begin{figure}
\centerline{\psfig{file=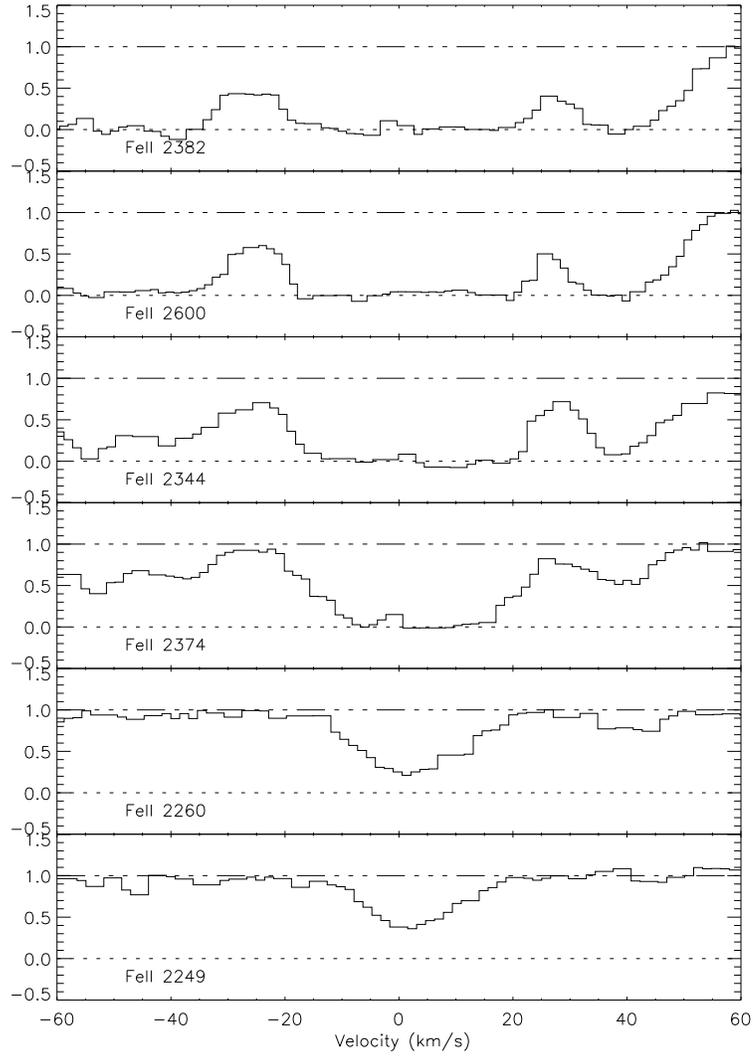,width=5.0in}}
\caption{Line profiles of \ion{Fe}{2} transitions plotted as a 
function of velocity relative to the systemic redshift of 
$z=1.54948$.
\label{fig:fs}}
\end{figure}

\begin{figure}
\centerline{\psfig{file=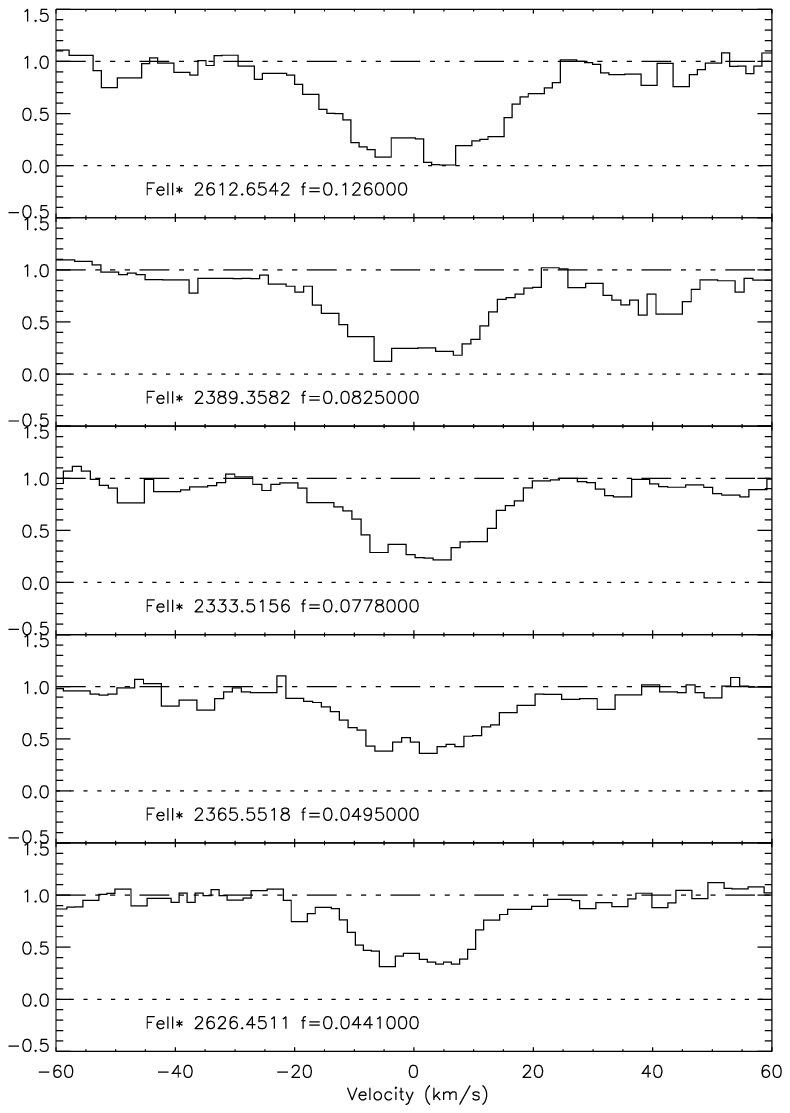,width=5.0in}}
\caption{Same as Figure~\ref{fig:fs} but for \ion{Fe}{2}*.
\label{fig:fsi}}
\end{figure}

\begin{figure}
\centerline{\psfig{file=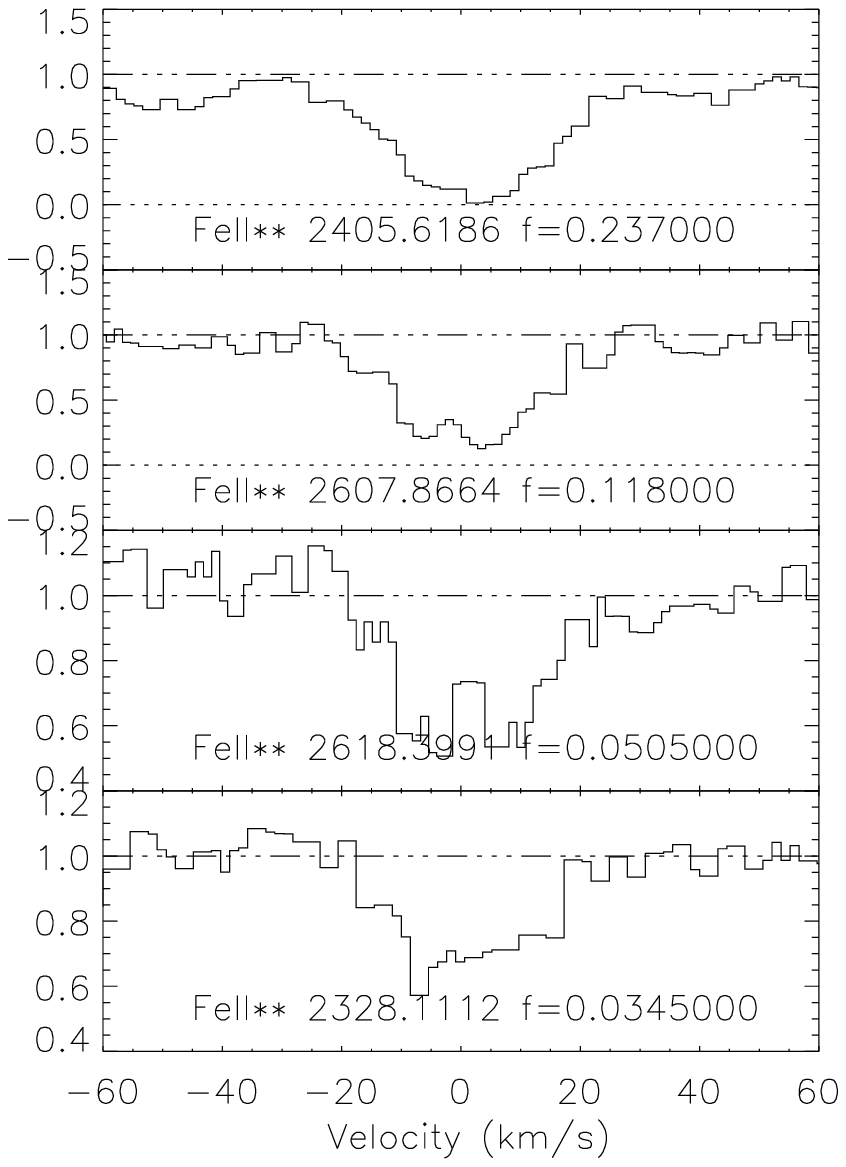,width=5.0in}}
\caption{Same as Figure~\ref{fig:fs} but for \ion{Fe}{2}**.
\label{fig:fsii}}
\end{figure}

\begin{figure}
\centerline{\psfig{file=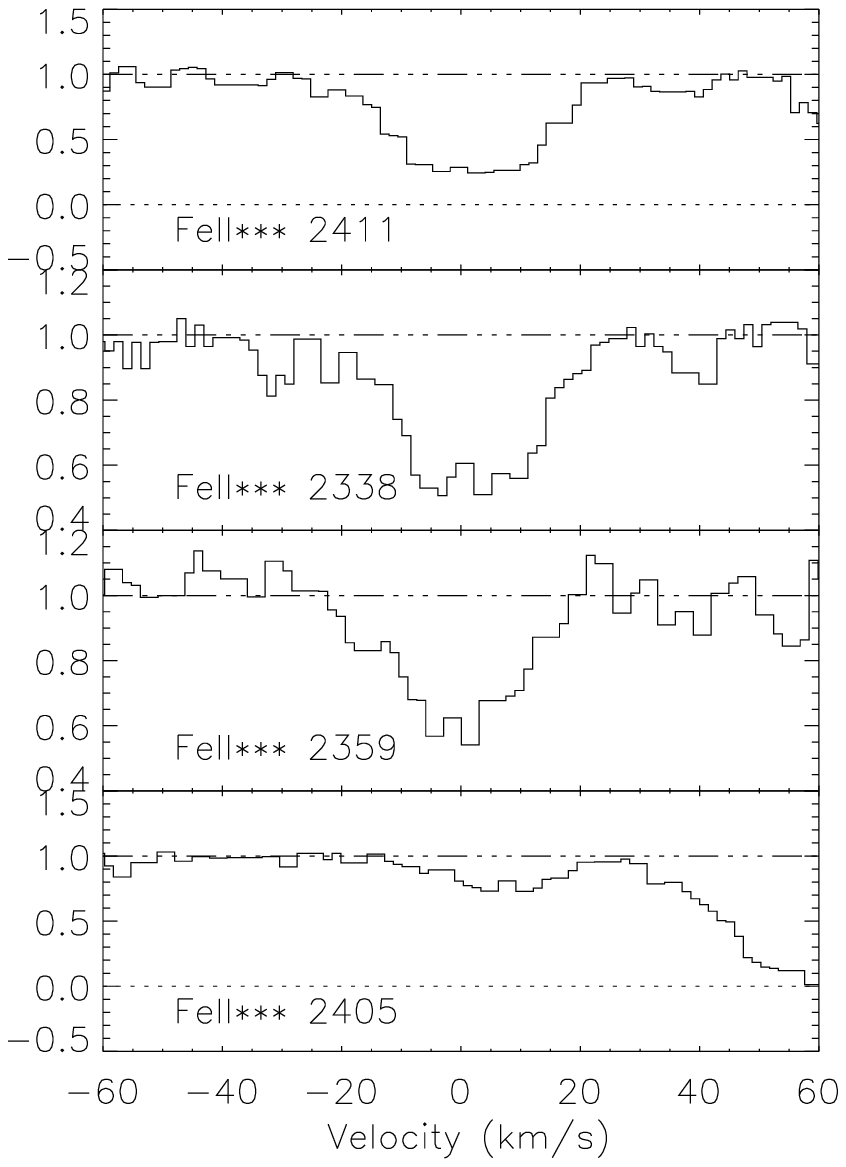,width=5.0in}}
\caption{Same as Figure~\ref{fig:fs} but for \ion{Fe}{2}***.
\label{fig:fsiii}}
\end{figure}

\begin{figure}
\centerline{\psfig{file=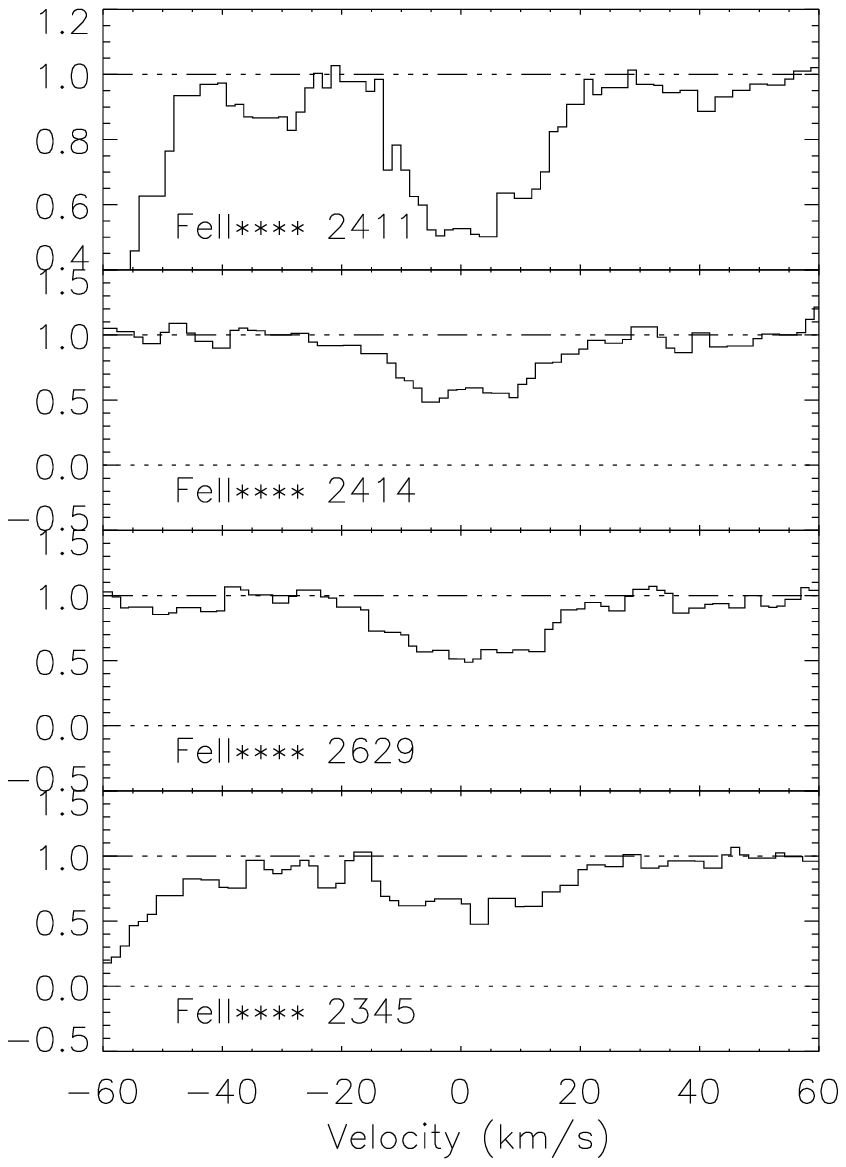,width=5.0in}}
\caption{Same as Figure~\ref{fig:fs} but for \ion{Fe}{2}****.
\label{fig:fsiiii}}
\end{figure}

In Figures~\ref{fig:fs}--\ref{fig:fsiiii} we plot the absorption
profiles of all transitions of the \ion{Fe}{2} ground and
fine-structure states as a function of velocity relative to the
systemic redshift of $z=1.54948$.  The line identifications, observed
wavelengths, and equivalent widths are summarized in
Table~\ref{tab:lines}.  Also included in the Table are the absorption
lines from the ground-state and fine-strcuture level of \ion{Si}{2}.
We detect absorption features from all four fine-structure levels of
the \ion{Fe}{2} $^3d^64s^6D$ level, (\ion{Fe}{2}* to \ion{Fe}{2}****),
correspoding to $J$-values of $7/2$, $5/2$, $3/2$, and $1/2$ (see also
\citealt{gcn4271}).  The absorption profiles exhibit a simple and
symmetric velocity structure extending from about $-30$ to $+30$ km
s$^{-1}$ and centered on the systemic velocity of the burst.  Strong
lines of the \ion{Fe}{2} ground-state level exhibit an overall
velocity range of about $-150$ to $+50$ km s$^{-1}$, but about $85\%$
of the gas is contained in the narrow velocity component
(\citealt{gcn4271}, Penprase et al. 2005).  We interpret the dominant
component to arise from the local environment of the burst based both
on the kinematic coincidence and the unusual detection of
fine-strcuture transitions.

The detection of the full range of \ion{Fe}{2} ground-level
fine-structure states is highly unusual, and to our knowledge
represents the first such example in an extra-galactic
source\footnotemark\footnotetext{Absorption from \ion{Fe}{2}* alone
was detected in the spectrum of GRB\,050730 \citep{cpb+05}.}.  In the
Milky Way, such transitions have been detected in the context of a few
dense environments such as in the circumstellar disks around $\beta$
Pictoris \citep{kb85} and 2 Andromedae \citep{cbn97}, and the HH 47A
bow shock \citep{hmt+99}.  Absorption from \ion{Si}{2}*, on the other
hand, is more prevalent and has been previously detected in several
GRB-DLAs \citep{sf04,vel+04,bpc+05}, as well as in the lensed Lyman
break galaxy MS 1512$-$cB58 \citep{prs+02}, but not in any QSO-DLAs
\citep{hwp05}.  This suggests that \ion{Si}{2}* may trace regions of
intense star formation, which tend to be missed in quasar sight lines
due to the cross-section selection effect or possibly the associated
dust extinction.

We measure the column densities of the various levels using the
apparent optical depth method \citep{ss91}.  This method has the
advantage that it makes no {\it a priori} assumptions about the
functional form of the velocity distribution, and at the same time it
incorporates information from a wide variety of lines with different
oscillator strengths, allowing a robust measure of the column density
in the line cores.  The column densities derived from this method by
integrating over the velocity range of $\pm 30$ km s$^{-1}$, as well
as the ratios of the fine-structure levels compared to the ground
state, are summarized in Table~\ref{tab:columns}.

\subsection{Collisional Excitation}
\label{sec:coll}

\begin{figure}
\centerline{\psfig{file=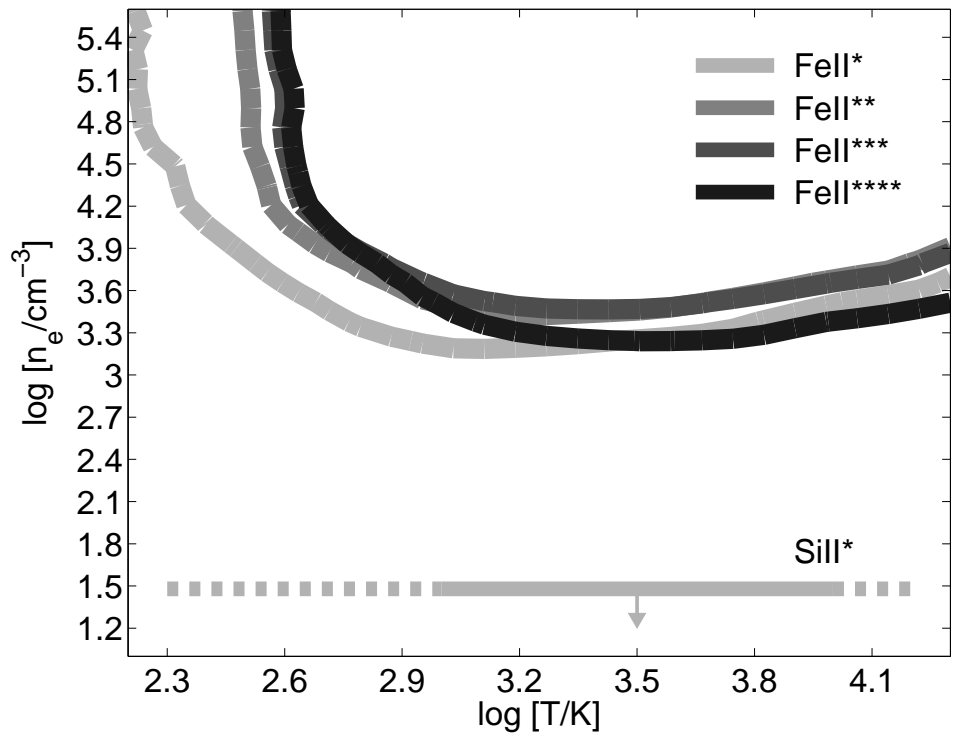,width=5.0in}}
\caption{Contours of temperature and electron density defined by the 
ratios of the four \ion{Fe}{2} fine-structure states to the ground
state, as well as the ratio of \ion{Si}{2}* to \ion{Si}{2}.  The 
physical conditions inferred from the \ion{Fe}{2} ratios are $T
\gtrsim 400$ K and $n_e\gtrsim 1.6\times 10^3$ cm$^{-3}$, with the
best-fit requiring $T\sim 10^3$ K and $n_e\sim 6\times 10^3$
cm$^{-3}$.  However, these conditions do not agree with the 
observed column density of \ion{Si}{2}* from which we derive a much
lower density.  The thickness of the lines designates a 0.15 dex
uncertainty in the ratio.
\label{fig:tne}}
\end{figure}

The excitation of the fine-structure levels can be achieved either by
collisions with ambient electrons, or by radiative pumping due to a
local radiation field.  Collisional excitation of \ion{Fe}{2} into its
fine-structure states has been studied by \citet{khb+88}, and more
recently by \citet{sv02}; the latter authors also investigate the
conditions for \ion{Si}{2}* collisional excitation.  In
Figure~\ref{fig:tne} we plot contours of the observed ratios of
\ion{Fe}{2} fine-structure columns relative to the ground-state in the
$T-n_e$ space, and find that the required electron density is at least
$2\times 10^3$ cm$^{-3}$.  Similarly, the detection of \ion{Fe}{2}****
lines requires a temperature of at least $400$ K.  Taking into account
all four ratios we find that the best-fit density and temperature are
${\rm log}\,n_e\approx 3.5-4$ and ${\rm log}\,T\approx 2.7-3.1$.  We
note, however, that the ratio of \ion{Fe}{2}* to \ion{Fe}{2} is about
a factor of two lower than expected based on the ratios of the three
other excited states.

A more significant discrepancy is observed in the ratio of the
\ion{Si}{2}* column relative to that of the ground state, $N({\rm
SiII^*})/ N({\rm SiII})\lesssim 0.06$.  This is an upper limit since
the single \ion{Si}{2}$\lambda 1808$ line detected in the spectrum is
saturated.  A comparison to other absorption features from which we
can determine the Doppler parameter $b$ (Penprase et al. 2005)
suggests that the ratio is about $0.022$.  Using the analysis of
\citet{sv02} we find that for the tempearure range of $10^3-10^4$ K,
the observed ratio indicates an electron density of only $\lesssim 20$
cm$^{-3}$.  Conversely, using the electron density inferred from the
\ion{Fe}{2} transitions we find an expected column density ratio for
\ion{Si}{2}* of about $0.5-1$, at least an order of magnitude larger
than detected.

The significant difference in physical conditions derived from the
\ion{Fe}{2} and \ion{Si}{2} fine-strcuture absorption suggests that
they either arise in different regions of the absorbing cloud, or that
a different mechanism, namely radiative pumping is at play.  We
consider the former possibility unlikely since there is no clear
reason why \ion{Fe}{2} and \ion{Si}{2} should be segregated within the
absorber, particularly in light of their similar velocity structure.

\subsection{Excitation by Radiative Pumping}
\label{sec:rad}

In the context of radiative pumping two possibilities exist.  First,
the fine-structure states can be directly excited by infra-red
radiation.  Successive levels starting with \ion{Fe}{2}* require
energies of $384.79$, $667.68$, $862.613$, and $977.053$ cm$^{-1}$,
respectively, for excitation.  The \ion{Si}{2}* level requires an
energy of $287.24$ cm$^{-1}$.  In the case of IR pumping the ratio 
of the fine-structure states relative to the ground state is given 
by (e.g., \citealt{sp01}),
\begin{equation} 
\frac{N({\rm X^*})}{N({\rm X})}=\frac{2n_\lambda}{1+n_\lambda}, 
\end{equation} 
where the number of photons at the appropriate excitation energy 
is, 
\begin{equation} 
n_\lambda=\frac{I_\nu \lambda^3}{8\pi hc}.  
\end{equation} 

Based on the observed column density ratios listed in
Table~\ref{tab:columns} and the energies of the various excitation
levels we infer the following specific intensities at the location 
of the \ion{Si}{2} and \ion{Fe}{2} absorbers: 
\begin{eqnarray}
I_\nu(8.62\times 10^{12}\,{\rm Hz})\lesssim 3.9\times 10^{-9}\,\, 
{\rm erg\, cm^{-2}\, s^{-1}\, Hz^{-1}} \\ 
I_\nu(1.15\times 10^{13}\,{\rm Hz})\approx 7.0\times 10^{-9}\,\, 
{\rm erg\, cm^{-2}\, s^{-1}\, Hz^{-1}} \\
I_\nu(2.00\times 10^{13}\,{\rm Hz})\approx 2.2\times 10^{-8}\,\, 
{\rm erg\, cm^{-2}\, s^{-1}\, Hz^{-1}} \\ 
I_\nu(2.59\times 10^{13}\,{\rm Hz})\approx 3.3\times 10^{-8}\,\, 
{\rm erg\, cm^{-2}\, s^{-1}\, Hz^{-1}} \\
I_\nu(2.93\times 10^{13}\,{\rm Hz})\approx 2.2\times 10^{-8}\,\, 
{\rm erg\, cm^{-2}\, s^{-1}\, Hz^{-1}} 
\end{eqnarray} 
These values roughly define a spectrum $F_\nu\propto \nu^{2.2}$ with
a possible turn-over at $\nu_0\gtrsim 6.4\times 10^{13}$ Hz.

A second possibility is indirect UV pumping in which decays from
short-lived excited UV levels populate the fine-structure levels in
addition to the ground level.  \citet{frs80} investigate this process
for the \ion{Si}{2} ion and find that UV pumping dominates over
collisional excitation as long as\footnotemark\footnotetext{The
expression assumes a flat spectrum which may not be directly
applicable if the UV pumping is due to the GRB UV flash or afterglow
emission, but we expect the difference to be minor.}  $\eta=8.5\times
10^{-14}n_e(T_e/10^4\,{\rm K})^{-1/2}(R_{\rm pc}/r_{17})^2I_\nu^{-1}
<1$.  A detailed investigation of this process for \ion{Fe}{2} has not
been performed to date.

\section{Star Formation versus GRB Excitation of the Fine-Structure 
Levels}
\label{sec:disc}

As discussed above, the mis-match in electron densities required to
explain the observerd ratios of \ion{Si}{2} and \ion{Fe}{2}
fine-strcuture levels relative to the ground state suggest that
collisional excitation is unlikely to be the source of the observed
fine-structure population.  In addition, we expect that if the
excitation is due to electron collisions produced as a result of the
GRB shock wave, then the velocity widths of the fine-structure lines
should be significantly broader than the observed $\pm 30$ km
s$^{-1}$.  Thus, excitation due to radiative pumping appears to be a
more likely scenario.  In this context it is instructive to consider
two sources for the IR or UV photons: star formation and the GRB
afterglow itself.

In the context of IR emission from stars, the inferred spectrum
$F_\nu\propto \nu^{\sim 2.2}$ with a possible turn-over at
$\nu_0\gtrsim 6.4\times 10^{13}$ Hz may be indicative of a thermal
spectrum with a peak of about $640$ K (or greater if the turnover is
not real), arising from dust re-radiation of absorbed star light.  The
much higher temperature compared to, for example, the IR emission from
Arp\,220 with $T\approx 30$ K, as well as the required large
luminosity, $L_\nu\approx 4\times 10^{30}(r/1\,{\rm pc})^2$ erg
s$^{-1}$ Hz$^{-1}$ (approaching $10^{10}$ L$_\odot$), indicate an
intense radiation field in the environment of \grb\ compared to those
in local star-forming regions.  The star formation rate required to
produce the inferred luminosity is in excess of 100 M$_\odot$
yr$^{-1}$ \citep{ken98}, and the high temperature requires a large
fraction of hot OB stars and a large dust content.  Perhaps a similar
analogue is the super star cluster in the dwarf galaxy NGC\,5253
($L\sim 10^9$ L$_\odot$), which contains $\sim 5000$ massive O stars
within a region of $\sim 1$ pc.  We note that if this is the correct
scenario, then it is likely that the GRB had to destroy dust along the
line of sight in order for the optical afterglow to be detected on
Earth.

In the context of UV radiative pumping the condition $\eta<1$ for
\ion{Si}{2}* excitation described in \S\ref{sec:rad} may be satisfied
in the vicinity ($\sim 0.1$ pc) of OB stars \citep{frs80}.  Thus, a
large concentration of such stars may lead to the observed
\ion{Si}{2}* column density.  The conditions for UV pumping of
\ion{Fe}{2} fine-structure states have not been investigated
previously and are beyond the scope of this paper, but we expect that
a more intense UV radiation field is required for excitation of
\ion{Fe}{2}**** compared to \ion{Si}{2}*.  Given that UV and IR
pumping require similar star formation conditions, it is likely that
both mechanisms operate within the star-forming region.

Finally, it is possible that the radiative pumping is due to the GRB
afterglow emission.  In this context we use the observed flux of the
afterglow, $F_\nu\sim 1-10$ mJy in the first hour, along with the IR
fluxes inferred in \S\ref{sec:rad}, to find that the burst-absorber
distance is about $10-20$ pc, which is reasonable for the size of a
star-forming region.  On the other hand, it is difficult to reconcile
the observed shape of the spectrum ($F_\nu\propto\nu^2$) with that of
a GRB afterglow ($F_\nu\propto \nu^{1/3}$), unless the excitation
takes place in different layers within the absorber.  In the case of
UV pumping by the GRB afterglow we find that for the observed flux
this process dominates over collisional excitation within a region of
about 10 pc unless the electron density exceeds $10^9$ cm$^{-3}$.

One possibility in the case of excitation by the GRB and afterglow
radiation is that the line strengths will change as a function of
time.  We repeated the column density calculation for each individual
exposure and found no significant evolution from $t\approx 1.2$ to 2.2
hours after the burst.

\section{Conclusions}
\label{sec:conc}

The high-resolution spectrum of \grb\ reveals a wide range of
absorption features from fine-structure states of \ion{Fe}{2} and
\ion{Si}{2} at $z=1.54948$.  We associate the absorption with the
local environment of the burst based on the fact that the excitation
of these fine-structure states requires large densities or intense
IR/UV radiation fields, which are not common in typical interstellar
environments.  This conclusion is also supported by the kinematic
coincidence with the burst systemic redshift.

A comparison of the conditions required to excite the \ion{Fe}{2} and
\ion{Si}{2} fine-structure levels indicates that collisional
excitation is not likely to be the dominant mechanism, unless the two
ions are segregated in different regions of the absorber.  A more
likely mechanuism is direct IR pumping or indirect UV pumping.  In
this scenario we find that if the radiation field is due to star
formation activity in the burst environment, then this requires a
large concentration of several thousand OB stars in a compact region
($\sim {\rm few}$ pc), perhaps reminiscent of the super star cluster
in the dwarf galaxy NGC\,5253.  If the UV and/or IR radiation are
supplied by the GRB, on the other hand, then the distance to the
absorber is likely to be about $10-20$ pc.  This is consistent with
calculations by \citet{dh02}, which indicate that within $\sim 3$ pc,
the burst will ionize its environment, leading to a very low column of
low-ionization states such as \ion{Fe}{2} and \ion{Si}{2}; outside of
this region it is likely that \ion{Fe}{2} and \ion{Si}{2} survive and
may be excited to fine-structure levels instead.

Clearly, detailed radiative transfer calculations, along with refined
calculations of \ion{Fe}{2} fine-structure excitation, are required in
order to distinguish between the different scenarios.  Still,
regardless of the exact details it is clear that GRB afterglows allow
us to directly probe the conditions within individual star forming
regions, and may therefore provide direct information on the mode of
star formation across a wide redshift range.  The absorption spectrum
of \grb\ seems to indicate that compact, dusty, and dense star forming
regions may be prevalent, at least in the context of GRB progenitor
formation.

\acknowledgements 
We thank Edward Jenkins and Bruce Draine for helpful discussion.  EB
is supported is supported by NASA through Hubble Fellowship grant
HST-01171.01 awarded by the Space Telescope Science Institute, which
is operated by AURA, Inc., for NASA under contract NAS 5-26555.


\clearpage
\begin{deluxetable}{llllll}
\tablecolumns{4}
\tabcolsep0.2in\footnotesize
\tablewidth{0pc}
\tablecaption{Line Identifications for \ion{Fe}{2} and \ion{Si}{2}
\label{tab:lines}}
\tablehead {
\colhead {$\lambda_{\rm obs}$}         &
\colhead {Line}                        &
\colhead {$f_{ij}$}                    &
\colhead {$W_0$}                       \\
\colhead {(\AA)}                       &
\colhead {(\AA)}                       &
\colhead {}                            &
\colhead {(\AA)}                       &
}
\startdata
4609.49 &  \ion{Si}{2}     1808.0130 & 0.00219 &  0.22796 \\
4632.22 &  \ion{Si}{2}*    1816.9285 & 0.00166 &  0.03723 \\
5736.02 &  \ion{Fe}{2}     2249.8768 & 0.00182 &  0.09490 \\
5763.81 &  \ion{Fe}{2}     2260.7805 & 0.00244 &  0.13672 \\
5935.47 &  \ion{Fe}{2}**   2328.1112 & 0.03450 &  0.07175 \\
5949.25 &  \ion{Fe}{2}*    2333.5156 & 0.07780 &  0.15565 \\
5962.53 &  \ion{Fe}{2}***  2338.7248 & 0.08970 &  0.10608 \\
5976.53 &  \ion{Fe}{2}     2344.2140 & 0.11400 &  0.37211 \\
5978.53 &  \ion{Fe}{2}**** 2345.0011 & 0.15300 &  0.10861 \\
6016.33 &  \ion{Fe}{2}***  2359.8278 & 0.06790 &  0.07552 \\
6030.93 &  \ion{Fe}{2}*    2365.5518 & 0.04950 &  0.13941 \\
6053.64 &  \ion{Fe}{2}     2374.4612 & 0.03130 &  0.29999 \\
6074.81 &  \ion{Fe}{2}     2382.7650 & 0.32000 &  0.41259 \\
6091.62 &  \ion{Fe}{2}*    2389.3582 & 0.08250 &  0.19033 \\
6131.92 &  \ion{Fe}{2}***  2405.1638 & 0.02600 &  0.05041 \\
6133.08 &  \ion{Fe}{2}**   2405.6186 & 0.23700 &  0.25156 \\
6147.44 &  \ion{Fe}{2}***  2411.2533 & 0.21000 &  0.18791 \\
6148.84 &  \ion{Fe}{2}**** 2411.8023 & 0.21000 &  0.10497 \\
6154.56 &  \ion{Fe}{2}**** 2414.0450 & 0.17500 &  0.10723 \\
6629.09 &  \ion{Fe}{2}     2600.1730 & 0.22390 &  0.44168 \\
6648.70 &  \ion{Fe}{2}**   2607.8664 & 0.11800 &  0.20213 \\
6660.91 &  \ion{Fe}{2}*    2612.6542 & 0.12600 &  0.25575 \\
6675.56 &  \ion{Fe}{2}**   2618.3991 & 0.05050 &  0.10022 \\
6685.89 &  \ion{Fe}{2}**** 2622.4518 & 0.05600 &  0.04794 \\
6696.08 &  \ion{Fe}{2}*    2626.4511 & 0.04410 &  0.15064 \\
6702.78 &  \ion{Fe}{2}**** 2629.0777 & 0.17300 &  0.12269 
\enddata
\tablecomments{Absorption features of \ion{Fe}{2} and \ion{Si}{2} 
ground and fine-structure states identified in the spectrum of \grb.
The columns are (left to right): (i) Observed wavelength, (ii) line
identification, (iii) oscillator strength \citep{pgw+03}, (iv)
redshift of the line, and (v) rest-frame equivalent width.}
\end{deluxetable}

\clearpage
\begin{deluxetable}{lll}
\tablecolumns{3}
\tabcolsep0.2in\footnotesize
\tablewidth{0pc}
\tablecaption{Column Densities of \ion{Fe}{2} and \ion{Si}{2} Ground
and Fine-Structure States
\label{tab:columns}}
\tablehead {
\colhead {Ion}		       &
\colhead {${\rm log}\,N$}      &
\colhead {${\rm log}\,(N^J/N)$} 
}
\startdata
\ion{Si}{2}     &  $>16.18$        & \nod     \\
\ion{Si}{2}*    &  $14.96\pm 0.10$ & $<-1.22$ \\
\ion{Fe}{2}     &  $15.21\pm 0.10$ & \nod     \\
\ion{Fe}{2}*    &  $13.89\pm 0.14$ & $-1.32$  \\
\ion{Fe}{2}**   &  $13.67\pm 0.17$ & $-1.54$  \\
\ion{Fe}{2}***  &  $13.52\pm 0.15$ & $-1.69$  \\
\ion{Fe}{2}**** &  $13.18\pm 0.20$ & $-2.03$  
\enddata
\tablecomments{Ionic column densities of \ion{Fe}{2} and \ion{Si}{2}
ground and fine-structure state.  The columns are (left to right):
(i) Ion, (ii) logarithm of the column density, and (iii) logarithm
of the ratio of fine-structure to ground state column density.}
\end{deluxetable}

\clearpage

\clearpage

\clearpage

\clearpage

\clearpage

\clearpage

\end{document}